**Plastic scintillation detectors for dose monitoring in digital breast tomosynthesis**

J. Antunes[1,2], J. Machado[1,2], L. Peralta[1,2], N. Matela[1,3]

[1]Departamento de Física da Faculdade de Ciências da Universidade de Lisboa
[2]Laboratório de Instrumentação e Física Experimental de Partículas
[3]Instituto de Biofísica e Engenharia Biomédica

ABSTRACT: Plastic scintillators detectors (PSDs) have been studied as dosimeters, since they provide a cost-effective alternative to conventional ionization chambers. Measurement and analysis of energy dependency were performed on a Siemens Mammomat tomograph for two different peak kilovoltages: 26 kV and 35 kV. Both PSD displayed good linearity for each energy considered and almost no energy dependence.

## Introduction

Breast cancer diagnose techniques, such as mammography and tomosynthesis, make use of X-ray beams with kilovoltage peaks ranging from 25 kV to 40 kV to obtain digital images of the organ.

Tomosynthesis constitutes a diagnose method that enables breast reconstruction from different radiographic projections. This technique uses ionizing radiation and so it is necessary to monitor the dose delivered to the patient avoiding over-exposure [1].

In recent years there has been increasing interest in the use of plastic scintillation detectors (PSD) due to its favorable characteristics compared with other systems [2]. In spite of this, there are no commercial systems in use for radiology application. Plastic scintillator-based dosimeters exhibit a linear relation between absorbed dose and produced signal, have good sensitivity and dose rate independence, possibility of direct readout and can be made quite small [3,4].

Diagnostic radiology typically uses X-rays beams between 25 and 150 kVp [5]. Studies of scintillator properties in this range have been done in recent years [6,2]. These works have shown that PSDs are good candidates for real-time dosimetry in limited energy ranges. Based on this concept a PSD with photomultiplier (PMT) readout was developed for radiology application. In particularly this device was tested on a beam delivered by a Tomosynthesis machine. The PSD consists of a plastic scintillator coupled to a clear PMMA optical fiber read by a photomultiplier. The signal produced by the photomultiplier is fed to a charge amplifier.

The aim of this study is to validate the device for tomosynthesis applications for two plastic scintillators, BC-404 and BCF-60. Previous prototype tests were conducted in the laboratory using 50 kVp X-ray beam followed by clinical validation on Mammomat Inspiration Tomograph from Siemens.

## Material and methods

*The dosimeter*

In this work two plastic scintillators (BC-404 and BCF-60), each of different bulk material and light emission properties [6] have been studied in view of their application as plastic scintillation detectors for low energy dosimetry.

Two dosimeters were built and evaluated, one using BC-404 blue-emitting scintillator and the other using BCF-60 green-emitting scintillator. The choice to use these scintillators was based on previous studies made on the scintillators performances [6]. The signal is transmitted through a clear PMMA optical fiber, 2 mm in diameter (ESKA SK-80 from Mitsubishi), to a PMT (Hamamatsu R647P). These fibers were chosen for their negligible fluorescence light yield. When present, this signal gives rise to what is known as stem effect, introducing an error in dose measurement. Our measurements using an optical fiber without a coupled scintillator on a 50 kVp X-ray beam showed a contribution to the total signal below 1%. The optical fiber was placed inside a black plastic jacket for isolation from room light. The PMT signal is amplified by charge amplifier read by a digital voltmeter. The charge amplifier was developed by our group and is based on the LCM6001 ultra-low input current integrated amplifier [7].

Dose measurements were made in a slab PMMA phantom $10.2 \times 10.8$ cm$^2$ on the side and 1.5 cm thick each slab. For measurements the phantom is assembled with 3 slabs in a total thickness of 4.5 cm. Special slabs with holes for holding the PSD or an ionization chamber were foreseen. For absolute dose measurements a Farmer ionization chamber (PTW TM30013) read by a PTW UNIDOS E electrometer was used. For this chamber the lower recommended energy by the manufacture is 30 keV, a typical value for this class of chambers. In order the use the same chamber for both beams and make comparison easier, the chamber was also used for the 26 kVp beam.

*The phantom*

A PMMA slab phantom was build for the measurements. Each slab is $10.2 \times 10.8 \times 1.5$ cm$^3$ and for each measurement three slabs were pile-up together and secured in place with plastic bolts (see figure 1). Two special slabs were manufactured with holes with the exact diameter, to accommodate the Farmer chamber or the PSD.

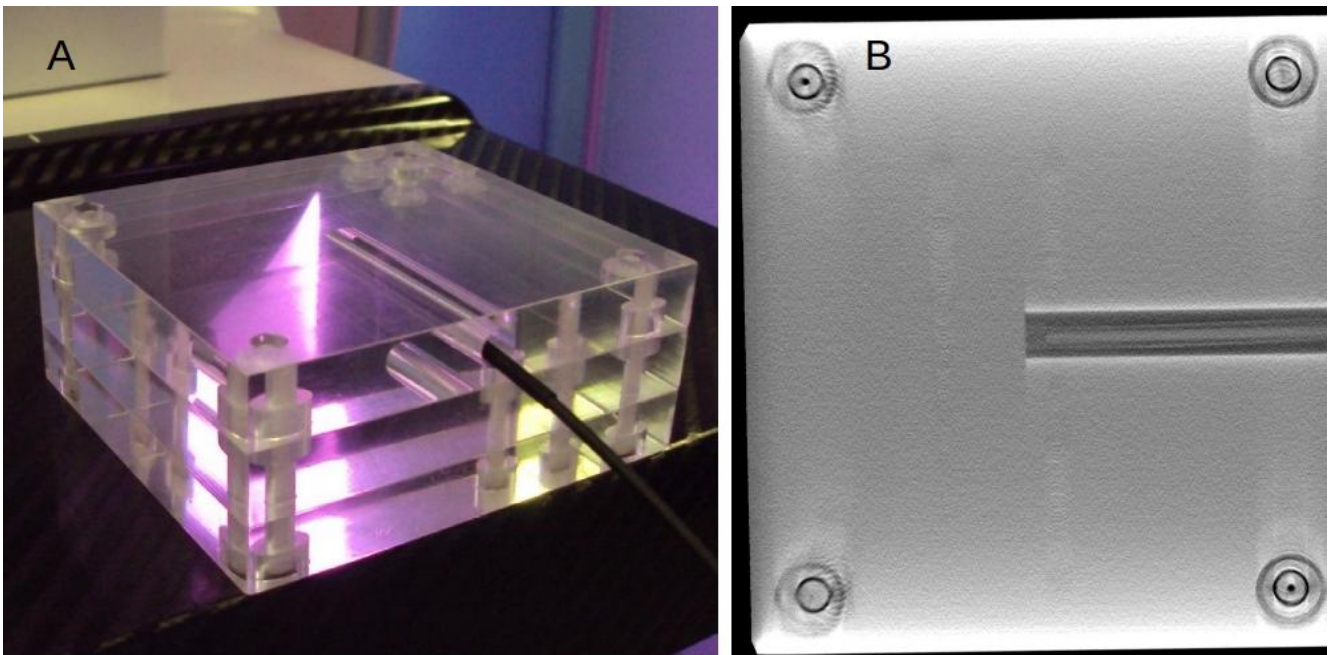

Figure 1: A) Detail of the phantom with the PSD inserted in the slab, B) Picture of the slab with hole for PSD obtained by the MAMMOMAT Inspiration System.

*The mammograph*

Clinical tests of the dosimeter were done at Hospital da Luz, Lisbon, at a MAMMOMAT Inspiration System (Siemens). During a tomoynthesis examination with MAMMOMAT Inspiration, a 50° sweep of the X-ray tube generates 25 images with a frame rate of up to 2 images per second (figure 2). In tomosynthesis acquisition mode, a W/Rh anode/filter combination is preprogrammed. The Rh filter is 50 µm thick and the produced X-ray beam exits through a 1 mm Be window. The phantom sits on top of the amorphous selenium (a-Se) detector at a focus to image-receiver distance of 65.55 cm. Our goal was to measure the dose delivered by the MAMMOMAT Inspiration System at different phantom depths with the signal obtained from a PSD. In the acquisition tests X-ray tube peak voltages of 26 and 35 kVp were used with total charge of 56 mA s in both cases. The total charge of 56 mA s is pre-set on the MAMMOMAT Inspiration System for standard image acquisition.

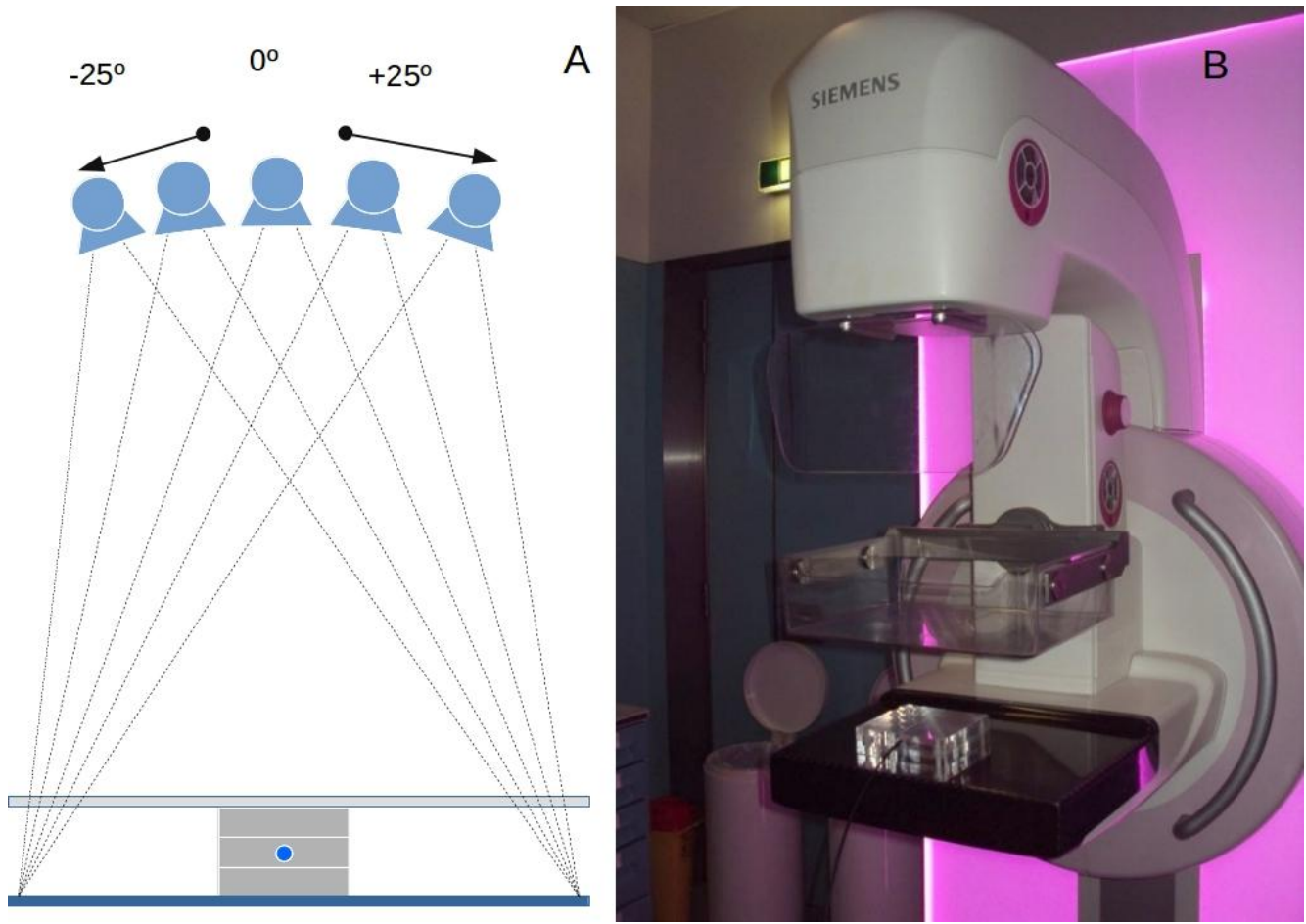

Figure 2: A) Schematic (not to scale) of the tomograph, compression plate, slab phantom and selenium plate. The X-ray tube makes a 50º sweep over the phantom; B) photo of the MAMMOMAT Inspiration System (Siemens) with slab phantom on top of selenium plate.

*Clinical tests*

Tomosynthesis is a technique that improves upon conventional mammography allowing a set of reconstructed planes acquired during a single motion of the X-ray tube [1]. Clinical tests were conducted at *Hospital da Luz* in Lisbon with a Mammomat Inspiration Tomograph from Siemens. The Mammomat Inspiration Tomograph X-ray tube has molybdenum and tungsten targets and molybdenum and rhodium filters [8]. Acquisitions were made at 26 and 35 kVp using the molybdenum anode and rhodium filter for a charge of 56 mA s.

The slab phantom was placed on the selenium detector with the compression plate on top of the phantom. The slabs holding PSD or ionization chamber could be moved to three different positions allowing three different dose measurements at depth 0.75, 2.25 and 3.75 cm.

**Results and discussion**

The results obtained in the clinical tests are presented in figure 3 for the PSD signal as a function of dose measured by the ionization chamber, and for two X-ray tube voltages. For each peak voltage the PSD or ionization chamber was placed in three slab positions (depth 0.75, 2.25 and 3.75 cm). Experimental uncertainties are lower than 2% both for PSD signal and ionization chamber. For each set of values a linear fit was performed to the data and the slope (sensitivity) extracted. Results are presented in table 1.

For the BC-404 scintillator the straight line slopes for 26 and 35 kV are identically within the experimental uncertainties. We thus conclude that the energy dependence of the PSD is small in the measurement range. Comparing the slopes for both scintillators we conclude that BC-404 has better performance using this PMT. R647P quantum efficiency peaks in the blue region, favoring the BC-404 blue-emitting scintillator over the BCF-60 green-emitting scintillator.

Table 1. Straight line slope from linear fit to PSDs signal as a function of dose inside a PMMA phantom when irradiated at Mammomat Inspiration Tomograph.

| BC-404: Slope (V/mGy) | | BCF-60: Slope (V/mGy) | |
| --- | --- | --- | --- |
| 26 kV | 35 kV | 26 kV | 35 kV |
| 0.81+-0.01 | 0.82+-0.02 | 0.281+-0.005 | 0.267+-0.006 |

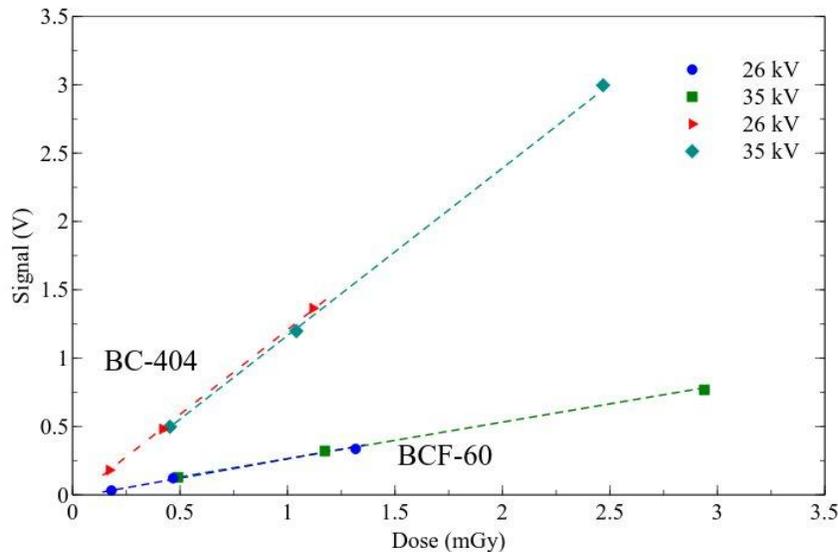

Figure 3. Signal from PSDs as a function of dose measured by an ionization chamber in a PMMA phantom irradiated at the Mammomat Inspiration Tomograph. Results are presented for BCF-60 and BC-404 scintillators and two X-ray tube peak voltages.

**Conclusion**

The adequacy of plastic scintillators to low energy dosimetry has been discussed in the literature [6, 9-10]. A dosimetry system based on plastic scintillator has been successfully tested for 26 and 35 kVp on a tomosynthesis system. The PSD energy dependence of the sensitivity is small in the measured range making it suitable for dose measurements with phantom in this radiological application. PSDs provide a low cost alternative for dose monitoring, and may be manufactured in a wide quality of shapes and dimensions.

**Acknowledgements**
This work was founded by Project PTDC/BBB-IMG/3310/2012 and project "QREN RAD4LIFE" CENTRO-07-ST24-FEDER-002007. We are thankful to *Hospital da Luz* in Lisbon for the opportunity of performing the clinical tests and to the LIP workshop in Coimbra for the manufacture of the phantom.